\documentclass[useAMS,usenatbib]{mn2e}
\usepackage{graphicx}

\def \ga {Geomagn. Aehron.}

\def \i3etm {IEEE Trans. Magn.}
\def \i3etns {IEEE Trans. Nucl. Sci.}
\def \i3etps {IEEE Trans. Plasma Sci.}

\def \jjap1 {Jpn. J. Appl. Phys., Part 1}
\def \jjap2 {Jpn. J. Appl. Phys., Part 2}

\def \pi3e {Proc. IEEE}

\def \sc {Sci} 

\def \sf {Spaceflight}

\def \usnos4 {U.S. Nav. Obs., Ser. 4}

\usepackage{times}
\usepackage[unicode,colorlinks,urlcolor=cyan,citecolor=blue,linkcolor=blue]{hyperref} 

\title[HST search for the transit of $\alpha$ Cen B\,b]{{\it Hubble Space Telescope} search for the transit of the Earth-mass exoplanet Alpha Centauri B\,b}

\author[Demory et al.]
{Brice-Olivier Demory\thanks{E-mail: bod21@cam.ac.uk}$^{1}$, David Ehrenreich$^2$, Didier Queloz$^1$, Sara Seager$^{3}$, \newauthor  Ronald Gilliland$^{4}$, William~J.~Chaplin$^{5,6}$, Charles Proffitt$^{7,8}$, Michael Gillon$^{9}$, \newauthor Maximilian N. G\"unther$^{1}$, Bj\"orn Benneke$^{10}$, Xavier Dumusque$^{11}$,  Christophe Lovis$^{2}$, \newauthor Francesco Pepe$^{2}$,  Damien S\'egransan$^{2}$, Amaury Triaud$^{12}$ \& St\'ephane Udry$^{2}$\\
$^1$Astrophysics Group, Cavendish Laboratory, J.J. Thomson Avenue, Cambridge CB3 0HE, UK. \\
$^2$Observatoire Astronomique de l'Universit\'e de Gen\`eve, 51 chemin des Maillettes, 1290 Versoix, Switzerland.\\
$^3$Department of Earth, Atmospheric and Planetary Sciences, Massachusetts Institute of Technology, 77 Massachusetts Ave., Cambridge, MA 02139, USA.\\
$^4$Department of Astronomy, and Center for Exoplanets and Habitable Worlds, The Pennsylvania State University, 525 Davey Lab, University Park, PA 16802, USA.\\
$^5$School of Physics and Astronomy, University of Birmingham, Edgbaston, Birmingham B15 2TT, UK.\\
$^6$Stellar Astrophysics Centre (SAC), Department of Physics and Astronomy, Aarhus University, Ny Munkegade 120, DK-8000 Aarhus C, Denmark.\\
$^7$Science Programs, Computer Sciences Corporation, 3700 San Martin Drive, Baltimore, MD 21218, USA.\\
$^8$Space Telescope Science Institute, 3700 San Martin Drive, Baltimore, MD 21218, USA.\\
$^9$Institut d'Astrophysique et de G\'eophysique, Universit\'e of Li\`ege, all\'ee du 6 Aout 17, B-4000 Li\`ege, Belgium.\\
$^{10}$Division of Geological and Planetary Sciences, California Institute of Technology, Pasadena, CA 91125, USA.\\
$^{11}$Harvard-Smithsonian Center for Astrophysics, 60 Garden Street, Cambridge, Massachusetts 02138, USA.\\
$^{12}$Massachusetts Institute of Technology, Kavli Institute for Astrophysics and Space Research, 77 Massachusetts Avenue, Cambridge, MA 02139, USA.}
 
\begin{document}

\date{Accepted 2015 March 25. Received 2015 March 16; in original form 2014 November 06}


\maketitle

\label{firstpage}

\begin{abstract}
Results from exoplanet surveys indicate that small planets (super-Earth size and below) are abundant in our Galaxy. However, little is known about their interiors and atmospheres. There is therefore a need to find small planets transiting bright stars, which would enable a detailed characterisation of this population of objects. We present the results of a search for the transit of the Earth-mass exoplanet Alpha Centauri B\,b with the Hubble Space Telescope (HST). We observed Alpha Centauri B twice in 2013 and 2014 for a total of 40 hours. We achieve a precision of 115 ppm per 6-s exposure time in a highly-saturated regime, which is found to be consistent across {\it HST} orbits. We rule out the transiting nature of Alpha Centauri B\,b with the orbital parameters published in the literature at 96.6\% confidence. We find in our data a single transit-like event that could be associated to another Earth-size planet in the system, on a longer period orbit. Our program demonstrates the ability of HST to obtain consistent, high-precision photometry of saturated stars over 26 hours of continuous observations.
\end{abstract}

\begin{keywords}
stars: individual: Alpha Centauri B -- techniques: photometric
\end{keywords}

\section{Introduction}

\label{intro}

Ground- and space-based observations have been providing a wealth of data on exoplanets covering a wide parameter regime. On one hand, dozens of detected hot Jupiters with high-SNR observations have expanded our knowledge of giant, irradiated exoplanets \citep[e.g.][]{Seager:2010a,Heng:2014b}. On the other hand lie the smaller, lower mass objects, of which terrestrial planets are a subset and about which very little is known. Both {\it Kepler} \citep{Batalha:2013} and radial-velocity surveys \citep{Howard:2010,Mayor:2011b} have been detecting an emerging population of these small, possibly terrestrial planets that appear to be ubiquitous. Very little is known about the interiors of small planets due to compositional degeneracies and their atmospheric properties are challenging to probe.  The search for and characterisation of terrestrial planets today represent the cutting edge of the exoplanet field. The harvest of results obtained on transiting giant planets shows the pathway to improve our knowledge of small planets: we have to find terrestrial exoplanets that transit bright stars, for which both masses and radii can be precisely measured. Despite the large number of Earth-size candidates found by {\it Kepler}, most of their host stars are too faint to precisely constrain the planetary masses using radial-velocity techniques. For most of the planets, we know only their sizes, not their masses or densities. However, radial-velocity surveys target close, bright stars. Thus, precisely monitoring the expected transit window of radial-velocity discovered planets \citep{Gillon:2011} could potentially result in the detection of a transit, the derivation of the planetary radius and advanced follow-up techniques. Successful cases are, among others, GJ\,436\,b \citep{Gillon:2007c}, 55\,Cnc\,e \citep{Demory:2011,Winn:2011a} and HD\,97658\,b \citep{Dragomir:2013a}.

The recent discovery of $\alpha$~Cen~B\,b with the HARPS spectrograph \citep{Dumusque:2012} is a significant achievement towards the discovery of an Earth twin, by demonstrating the ability of the radial-velocity technique to detect Earth-mass companions. With a Doppler semi-amplitude of only 0.51$\pm$0.04 m/s, $\alpha$~Cen~B\,b is the lowest RV-amplitude planet found as of today. Four years of data and a careful analysis were necessary for the authors to characterise the stellar oscillation modes, rotational-induced and long-term activity as well as the binary orbital motion. These critical steps revealed the planetary signal and underlined the need of simultaneously fitting the stellar and planetary signals to push the precision of radial-velocity surveys \citep[e.g.][]{Boisse:2010}. The complexity of the analysis, coupled to the weak planetary signal could be seen as the main challenge of this major discovery \citep{Hatzes:2013a}, calling for an independent confirmation. The motivation of the present study was to carry out a photometric confirmation of this planet with the Hubble Space Telescope (HST). With a geometric probability of 9.5\%, the detection of a transit of $\alpha$ Cen B\,b would be first and foremost a clear confirmation that the planet is real, and would represent an opportunity to precisely constrain the density of an Earth-mass planet.

\section{Observations and Data Reduction}
\label{datared}

We first observed $\alpha$~Cen~B (V = 1.33, B-V = 0.90, K1V) quasi-continuously for 16 orbits (26 hours) on July 7\---8 2013 with the Space Telescope Imaging Spectrograph (STIS). This long monitoring sequence allowed us to cover $\sim$96\% of the transit window allowed by the radial-velocity orbital solution of \citet{Dumusque:2012}. 

\begin{figure}
\begin{centering}
\includegraphics[width=0.5\textwidth]{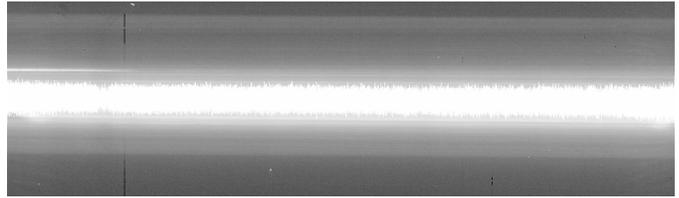}
\caption{{\bf 2D raw extracted spectrum.} This spectrum is a sample from our program PID13180. The degree of saturation as well as scattered light from the star are both clearly visible.
\label{fig:spec}}
\end{centering}
\end{figure}

The brighter $\alpha$~Cen~A (V = -0.01, B-V = 0.69, G2 V) was 4.5" away from the B component in July 2013. The small apparent separation and respective brightnesses of both components of the $\alpha$~Cen~system required tailored observational settings. In particular, the stellar brightness forced us to saturate the exposures; not saturating them would have meant setting short exposure times commensurable with the detector shutter closing time, yielding high uncertainties on the actual values of the exposure time (shutter-timing jitter). While previous studies found that saturated spectra do not preclude achieving high signal-to-noise ratio photometry on bright stars \citep{Gilliland:1999,Gilliland:1999a}, saturating $\alpha$~Cen~B implies that $\alpha$~Cen~A would saturate even more, making it impossible to have both stars in the spectrograph slit without having their blending charges overlapping. To avoid this situation, we positioned the spectrograph slit on $\alpha$~Cen~B and requested a range of telescope roll angles such as $\alpha$~Cen~A would be masked out and aligned in the dispersion direction. This configuration indeed minimises any contamination by $\alpha$~Cen~A and the potential impact of its 45-degree diffraction spikes. HST, however, does not actually roll on request but presents varying roll angle as a function of time. In this case, it was possible to schedule observations of $\alpha$~Cen~B in the HST continuous viewing zone (free of Earth occultations) while committing with the selected range of telescope roll angles. The target acquisition was first performed on the brighter $\alpha$~Cen~A before offsetting to $\alpha$~Cen~B. The target was then carefully centred in the $52''\times0.05''$ long slit. A wider slit ($52''\times2''$) was finally used to minimise slit losses for science exposures. We show for illustration purposes one spectrum of our program in Figure~\ref{fig:spec}.

We then re-observed $\alpha$~Cen~B for 9 orbits on July 27\---28 2014 to confirm a transit-shaped event detected in the July 2013 dataset. We provide details about our data reduction and analysis in the following section.

\subsection{July 2013 dataset}

We observed $\alpha$~Cen~B using STIS/CCD with the G750M grating and an exposure time of 6 s. We use the setting 6094, which has a wavelength domain ranging from 5812 to 6380\AA and a resolution of $\sim$5500.

\subsubsection{Data Reduction}

\begin{figure*}
\begin{centering}
\includegraphics[width=\textwidth]{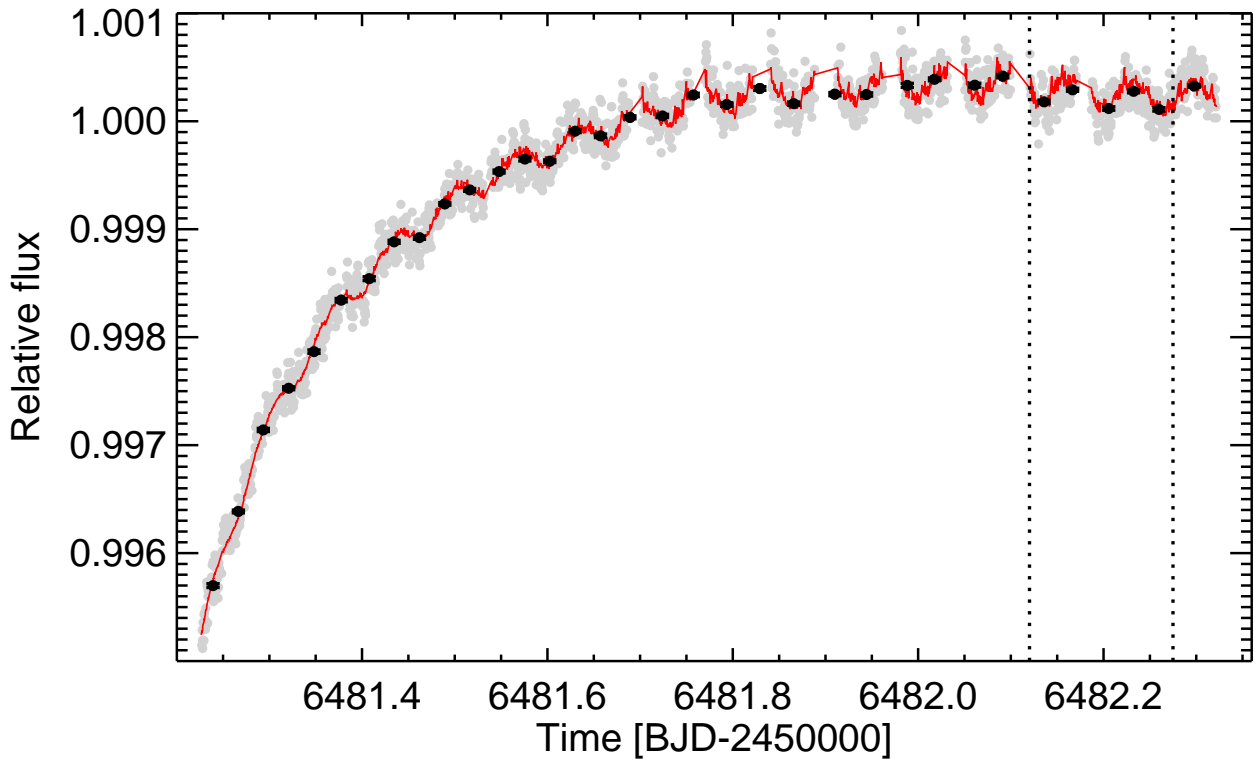}
\caption{{\bf Raw HST/STIS photometry.} The single-baseline best-fit MCMC model is superimposed in red. The dotted lines indicate the location and width of the transit pattern described in Section~\ref{analysis}.
\label{fig:rawp1}}
\end{centering}
\end{figure*}

The starting point of our data reduction consists of 11 flat-fielded science files (\textsc{flt}) that are the output of the STIS calibration pipeline \textsc{calstis}. These data are available on MAST under program ID 13180. Each \textsc{flt} file is a FITS data cube that contains a 2D spectrum for each exposure (2087 in total). All spectra are 1024 pixels in the dispersion direction and 300 pixels in the cross-dispersion direction. Since $\alpha$~Cen~B saturates the detector in the adopted 6-s exposure time, this larger acquisition window allowed us to integrate those columns that show bleeding up to $\sim$60 pixels along the cross-dispersion axis. We identify and discard cosmic ray hits and bad pixels on each calibrated frame. The HJD mid-exposure time is obtained from the file headers and converted to BJD$_{TDB}$ using existing routines \citep{Eastman:2010a}.  We use custom-built procedures to perform two one-dimensional spectral extractions using rectangular apertures with 120 and 180 pixels in the cross-dispersion axis centred on the spectrum. We perform the extraction with two different apertures initially to ensure that the photometry is insensitive to the aperture size. We apply these extractions to the 11 \textsc{flt} data cubes. We hold both the aperture size and position fixed for all spectral frames. We finally sum the spectra to build a high S/N ratio white light-curve for the photometric analysis. We collect $\sim$6 10$^9$ e$^{-}$ per 6-s exposure time. We find no evidence of contamination from the A component. Visual inspection of the raw photometric time-series shows a $\sim$0.4\% increase in flux for the first 0.6 days combined to a sinusoidal signal that matches HST's orbital period. The raw photometry is shown in Figure~\ref{fig:rawp1}.

\begin{figure*}
\begin{centering}
\includegraphics[width=\textwidth]{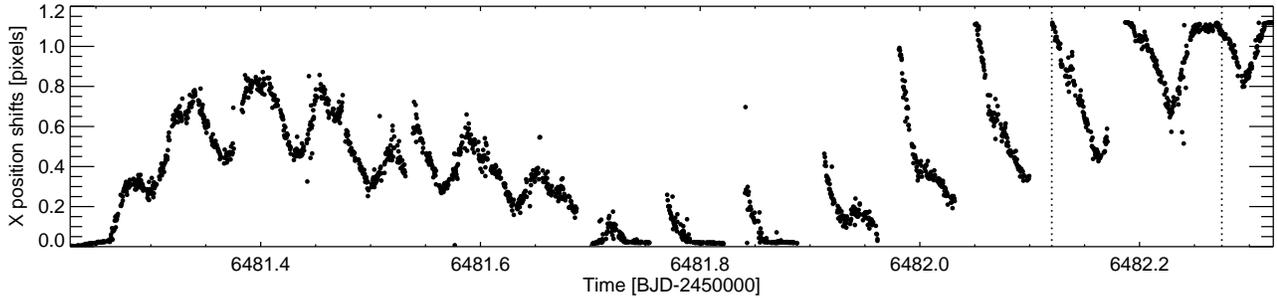}
\caption{{\bf Spectral shifts in X.} Displacement of the spectrum in the dispersion direction as calibrated using the Sodium doublet. The dotted lines indicate the location and width of the transit pattern described in Section~\ref{analysis}.
\label{fig:posxp1}}
\end{centering}
\end{figure*}

\begin{figure*}
\begin{centering}
\includegraphics[width=\textwidth]{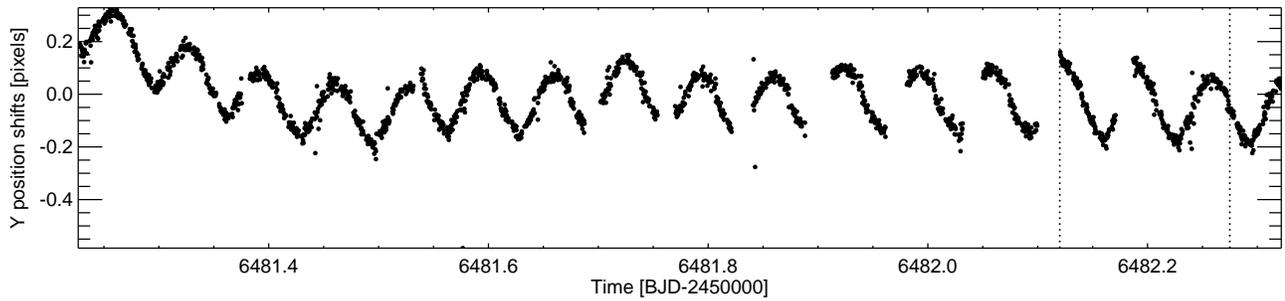}
\caption{{\bf Spectral shifts in Y.} Displacement of the spectrum in the cross-dispersion direction. The dotted lines indicate the location and width of the transit pattern described in Section~\ref{analysis}.
\label{fig:posyp1}}
\end{centering}
\end{figure*}

Tracking the spectral shifts on the detector is rendered difficult because of the level of saturation. To monitor displacements in the dispersion direction, we locate and fit the Na doublet lines 40 pixels from the spectral intensity centreline in the cross-dispersion direction to avoid performing the fitting in the saturated region. In the cross-dispersion direction, a section of the spectrum (at a given wavelength) can be approximated by a boxcar function, which is 1 for saturated pixels and 0 elsewhere, that we fit in seven locations along the dispersion direction of the spectrum. We use these multiple determinations to estimate an average value for the spectrum centre for each frame. We respectively detect drifts of 0.33 pixels RMS in the dispersion direction and 0.11 pixels RMS in the cross-dispersion direction. We show in Figures~\ref{fig:posxp1} and \ref{fig:posyp1} the spectral shifts in both directions during the entire visit.

We measure the median background level in two 10$\times$1022 pixel regions located at the subarray top edge and in another area of the same size at the bottom edge of the spectrum. Time-series extracted from both regions have a slightly different shape but their Fourier spectra have the same characteristic frequencies. We find a background level of about 54 e$^{-}$/pixel during the course of our observations, which is likely due to scattered light from $\alpha$~Cen~B. Examination of the background time-series reveals a clear periodicity at HST's orbital period as expected, that accounts for most of the measured background RMS at the 1\% level. The residual RMS for the background time-series is 0.02\%. We assess the incidence of the background on the time-series by subtracting its median value from each pixel in the aperture window. We find no effect on the photometry.

In the next step we focus on the impact of the flat-field correction on the photometric time-series. Calibrated 2D spectra available on MAST\footnote{http://mast.stsci.edu} have been reduced using the \textsc{x6417097o-pfl.fits} pixel-to-pixel flat file, whose RMS across the full frame is $\sim$1\%. As previously discussed \citep{Gilliland:1999}, the STIS CCD has a very flat response across its wavelength range. The flat field noise per pixel drift is estimated by dividing its RMS by the square root of the total number of pixels over which the incoming flux is spread on the detector. The stellar flux is spread over $\sim$60 pixels in the cross-dispersion direction and 1022 pixels in the dispersion axis (the two extreme columns are trimmed). The noise induced by the imperfect pixel-to-pixel flat is thus $\sim$38 ppm per pixel shift, which is below the measured individual errors of 115 ppm per 6-s exposure. As a second check, we also recalibrate the data without the flat-field correction, which results in photometric time-series that are indistinguishable from the ones including the pixel-to-pixel sensitivity correction.

We measure the overscan level in a fixed 12$\times$120 pixel region located at the extreme left of the \textsc{raw} files (as both overscan regions are trimmed during the 2D spectral calibration). Overscan time-series exhibit an asymptotic decrease reaching a stable level 0.6 days after the start of the observations. This trend is combined with a higher frequency variation matching HST's orbital phase 96-min duration. 

For each spectrum we record the timestamp, normalised flux, photometric error, spectral shift in the dispersion and cross-dispersion directions, median background and overscan values as well as HST's orbital phase at which each spectrum has been acquired.

\subsubsection{Photometric Analysis}
\label{analysis}

We use the MCMC algorithm implementation already presented in the literature \citep[e.g.][]{Gillon:2012a}. Inputs to the MCMC are the photometric time-series obtained during the data reduction described above. Photometric baseline model coefficients used for detrending are determined at each step of the MCMC procedure using a singular value decomposition method \citep{Press:1992}. The resulting coefficients are then used to correct the raw photometric lightcurves.

The baseline model for the full, 26-hr long time-series (see Figure~\ref{fig:rawp1}) consists of a second order logarithmic ramp \citep[e.g.][]{Knutson:2008, Demory:2011}, combined with a linear function of time and a fourth-order polynomial of {\it HST}'s orbital phase. The choice of the function best fitting the HST orbital phase effect, usually attributed to telescope breathing, is commonly based on half HST orbits \citep[e.g.][]{Sing:2009b,Huitson:2012,Evans:2013} because of the Earth occultations. Here, we are able to establish this function using the whole HST orbit and confirm that it is well approximated by a 4th-order polynomial. The logarithmic ramp is necessary to reproduce the increase in flux seen by visual inspection of the time-series. 

\begin{figure*}
\begin{centering}
\includegraphics[width=\textwidth]{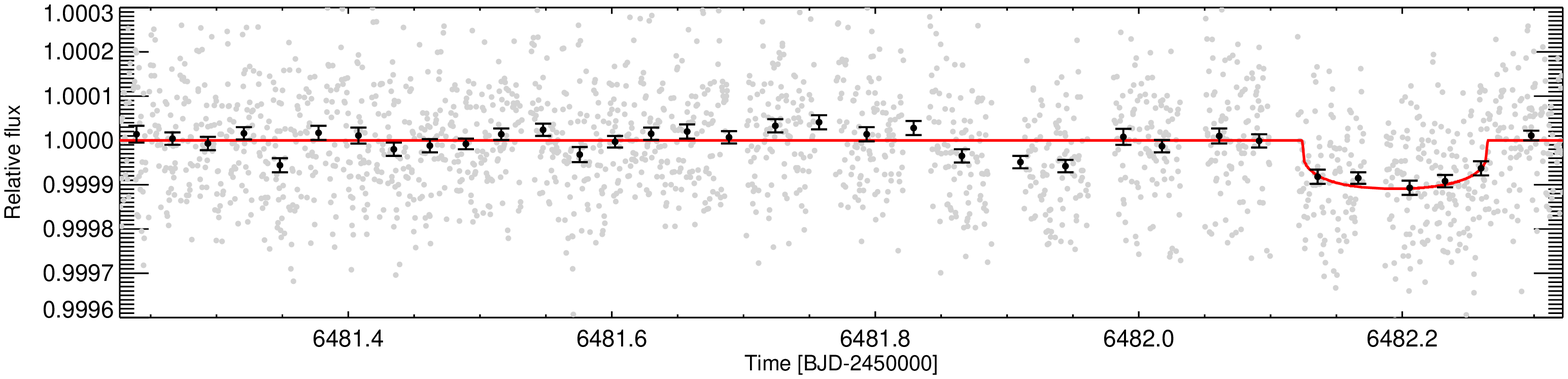}
\caption{{\bf Detrended photometry \--- unsegmented.} Data are binned per 45 minutes. The best fit model is superimposed in red.
\label{fig:unsegp1}}
\end{centering}
\end{figure*}

\begin{figure*}
\begin{centering}
\includegraphics[width=\textwidth]{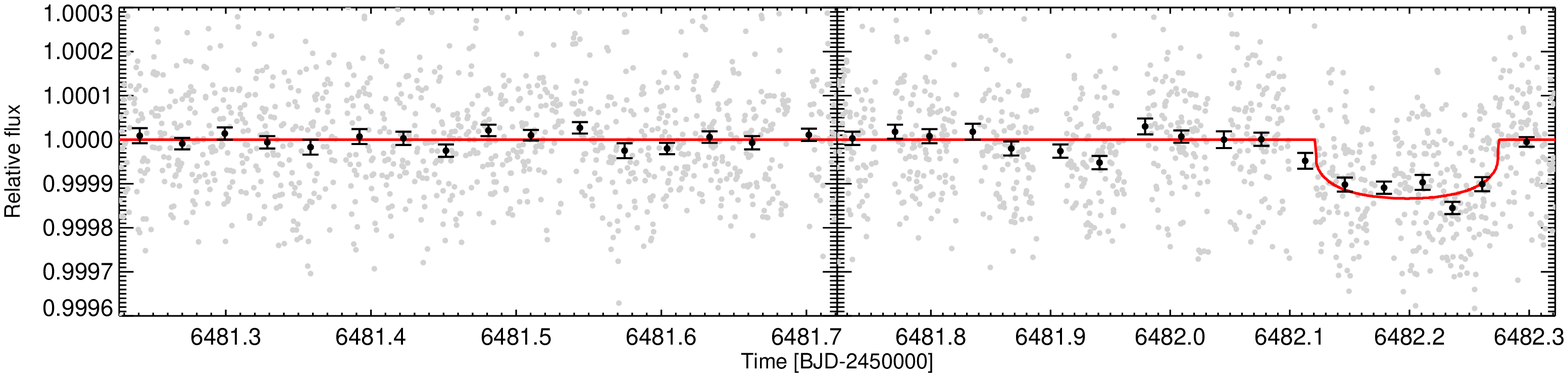}
\caption{{\bf Detrended photometry \--- 2 segments.} Data are binned per 45 minutes. The best fit model is superimposed in red.
\label{fig:2segp1}}
\end{centering}
\end{figure*}

\begin{figure*}
\begin{centering}
\includegraphics[width=\textwidth]{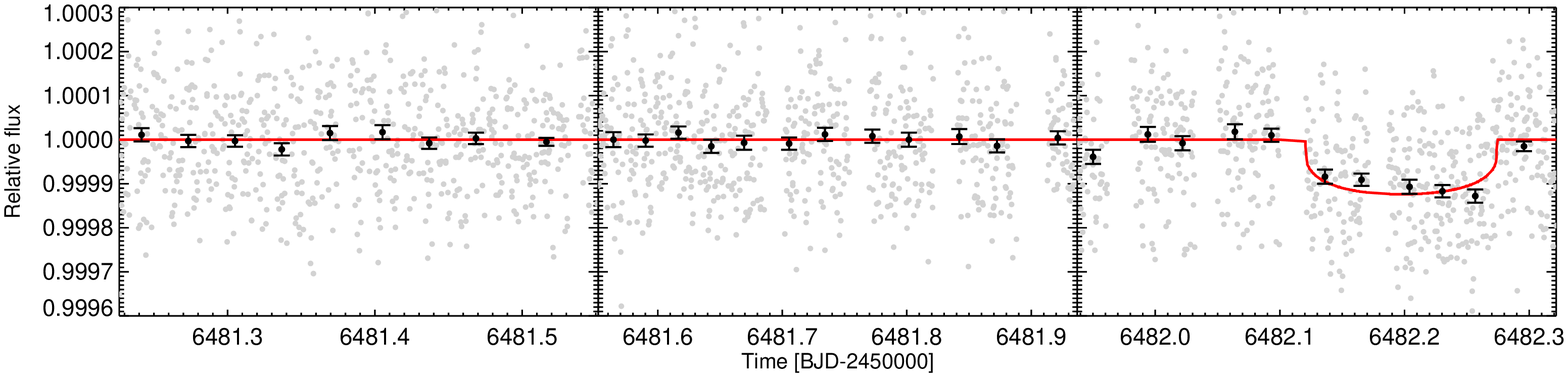}
\caption{{\bf Detrended photometry \--- 3 segments.} Data are binned per 45 minutes. The best fit model is superimposed in red.
\label{fig:3segp1}}
\end{centering}
\end{figure*}

\begin{figure*}
\begin{centering}
\includegraphics[width=\textwidth]{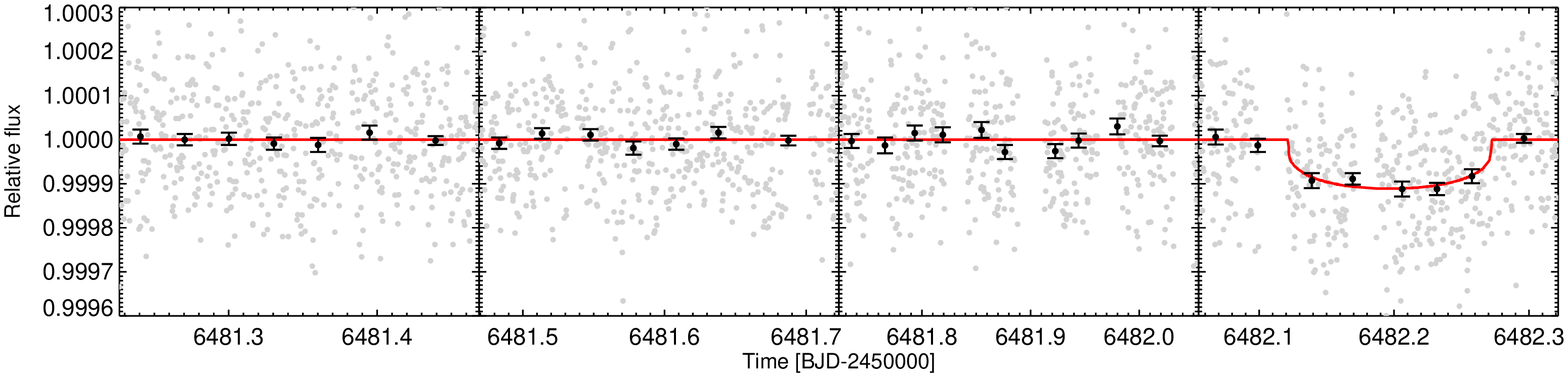}
\caption{{\bf Detrended photometry \--- 4 segments.} Data are binned per 45 minutes. The best fit model is superimposed in red.
\label{fig:4segp1}}
\end{centering}
\end{figure*}

This basic detrending reveals correlated structures in the photometry suggesting an imperfect correction. It is indeed unlikely that 26-hr of {\it HST} continuous observations can be accurately modelled with a single baseline model (constrained by fixed coefficients) as simple as the one described above  (see Figure~\ref{fig:unsegp1}). Since we do not know a priori if a transit exists in the data and what its location could be, we analyse the time-series by separating them in different segments spanning four to eight orbits, similar to other {\it HST} exoplanet programs. All segments are analysed in the same MCMC fit, with their own baseline model coefficients, but retaining the same functional form (orbital phase dependent polynomial and linear trend). We first split our photometric time-series into two segments. The logarithmic ramp term is needed for the first segment only. We then split our data into three segments. Since the ramp seen on the whole photometry affects the first 0.6 days of our data, the baseline models for the first two segments include a logarithmic ramp model. The time-series are finally split into four segments. For this 4-segment MCMC fit, only the first two files have a baseline including a ramp model. The results for these 3 MCMC analyses are shown in Figures~\ref{fig:2segp1}, \ref{fig:3segp1} and \ref{fig:4segp1}. We find that splitting the photometry in segments significantly decreases the Bayesian Information Criterion \citep[BIC,][]{Schwarz:1978} as compared to a unique fit to the whole photometry, despite the significant increase in free parameters. The behaviour of HST/STIS combined systematics slightly evolve from one orbit to another, making it difficult to model 16 orbits with a unique set of baseline coefficients. As expected, most structures found in the full photometry fit have vanished, but one, centred on $T_0 = 2456482.195$ BJD with a $\sim$0.15-day duration (first-to-fourth contact). This transit-like structure has a location, depth and duration that remains consistent whether the data are unsegmented or split in 2, 3 or 4 segments (see Figures~\ref{fig:unsegp1},\ref{fig:2segp1}, \ref{fig:3segp1} and \ref{fig:4segp1}). We notice another structure in the time-series with a 0.2 to 3-$\sigma$ significance (depending on the number of segments) and located at $T_0 = 2456481.9$ BJD. However, we find the duration and depth of this structure to change significantly with the baseline model and the number of segments used in the analysis, which suggests an instrumental origin. For comparison purposes, the transit-like pattern centred on $T_0 = 2456482.195$ BJD is detected at the 7.5-$\sigma$ level. We show in Table~\ref{tab:mcmcres} the MCMC fit results obtained for each analysis (unsegmented and segmented by 2, 3 and 4 parts).

\begin{table*}
\begin{tabular}{lllll} 
   & unsegmented & 2-segment & 3-segment & 4-segment   \\ \hline
Transit depth [ppm] & $91.3_{-18}^{+20}$ & $97.3_{-15}^{+12}$ & $90.9_{-13}^{+12}$ & $90.2_{-12}^{+12}$\\   
Transit duration [days] & $0.141_{-0.13}^{+0.29}$ & $0.151_{  -0.017}^{+0.016}$ & $0.153_{-0.011}^{+0.017}$ & $0.150_{-0.007} ^{+0.007}$   \\
Transit T0 [HJD] & $6482.194_{-0.006}^{0.008}$ & $6482.196_{-0.006}^{+0.005}$ & $6482.197_{-0.005}^{+0.004}$ & $6482.195_{-0.004}^{+0.003}$   \\
BIC & $1995$ & $1923$ & $1965$ & $1975$   \\
\end{tabular} 
\caption{{\bf Transit candidate fit results.} MCMC fit results obtained for each analysis (unsegmented and segmented by 2, 3 and 4 parts). Values are the median of the posterior distributions for each perturbed parameter indicated on the left column, along with the 1-$\sigma$ credible intervals. BIC values are calculated for the combined baseline+transit model.
\label{tab:mcmcres}}
\end{table*}

We then investigate whether the inclusion of ancillary parameters measured during the data reduction helps in improving the fit. We find that including the median overscan value as a linear coefficient in the baseline function improves the BIC by 3\%. The inclusion of linear combinations of the spectrum drifts in the dispersion and cross-dispersion directions increases the BIC by 5\%. Our final baseline model therefore consists of HST's orbital phase's fourth order polynomial, a time- and overscan-dependent linear trends. Adding higher-order terms to these functions does not improve the BIC.

\begin{figure*}
\begin{centering}
\includegraphics[width=\textwidth]{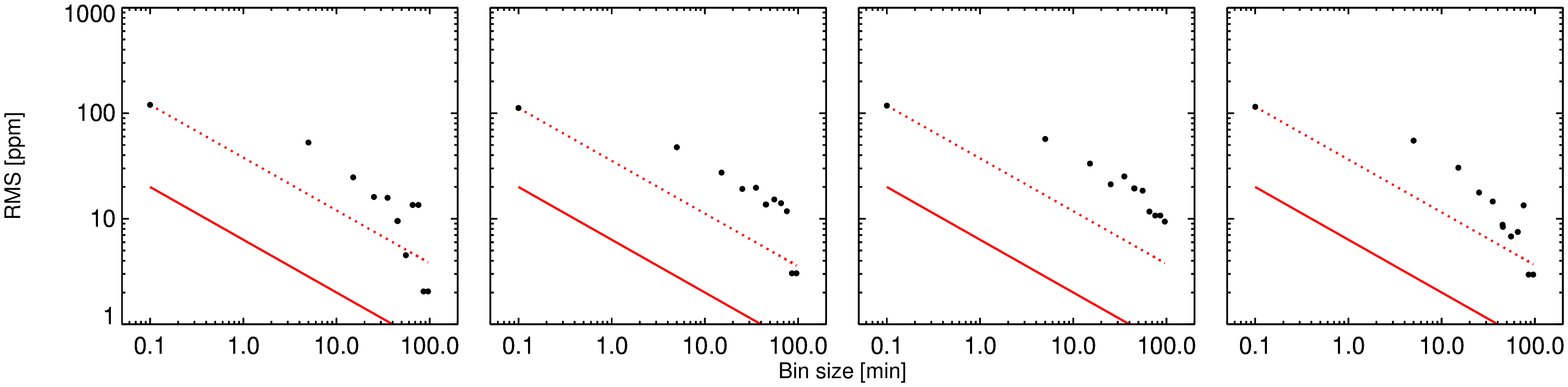}
\caption{{\bf RMS of binned residuals versus bin-size.} The first panel corresponds to the first segment, the fourth panel to the last segment. The theoretical photon noise is shown as a solid red line. The photometric RMS is indicated vs. the time-series bin size (black points). The expected decrease in Poisson noise normalised to an individual exposure (6s) is shown as a red dotted line. 
\label{fig:rmsp1}}
\end{centering}
\end{figure*}

We find a consistent exposure-to-exposure photometric precision of 110 to 120 ppm depending on the segment, significantly above the Poisson-limited precision of 20 ppm per exposure. The origin of this large discrepancy is not known but could be due to a larger than expected high-frequency instrumental noise due to shutter-timing jitter or could also have a contribution from stellar granulation or pulsations. We measure low contamination from correlated noise \citep{Pont:2006b} that we estimate  between 22\% and 34\% following \citet{Gillon:2010a} depending on the segment. Figure~\ref{fig:rmsp1} shows the actual decrease of the photometric RMS with the time-series bin size (black points). The theoretical photon noise is shown as a solid red line. We also indicate the expected decrease in Poisson noise normalised to the individual exposure time (in red dotted line). Despite the larger discrepancy between Poisson noise and our actual precision, Figure~\ref{fig:rmsp1} shows that this extra noise component has mostly a white behaviour that does not affect our science goal (transit detection). The improvement in precision is even accentuated for the larger bin sizes (60 min and more), meaning that a large part of the extra noise noticed at higher frequencies reduces significantly for larger time bins, which correspond to typical transit durations. We adjust our individual error bars taking into account these extra white and red noise components within the MCMC framework.

\begin{figure*}
\begin{centering}
\includegraphics[width=0.5\textwidth]{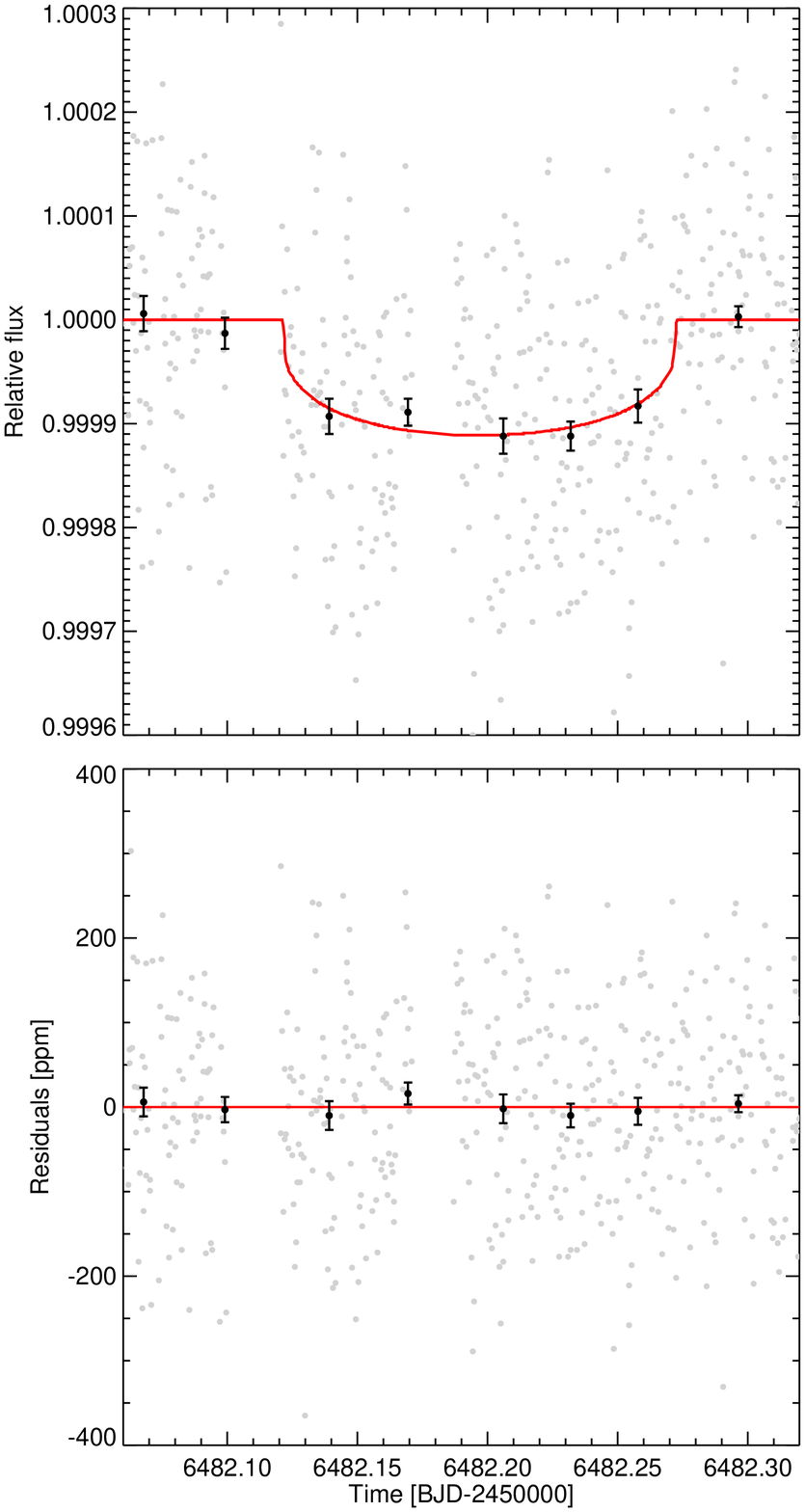}
\caption{{\bf Corrected HST/STIS corrected light-curve centered on the transit signal.} Data are binned per 45 minutes. Unbinned datapoints are shown in grey. Our best-fit MCMC model is superimposed in red.
\label{fig:transp1}}
\end{centering}
\end{figure*}

In the following we focus on the observed transit structure centred on 2456482.2 BJD. The second half of the transit structure benefits from {\it HST} continuous monitoring with no interruptions due to the SAA, which enabled the continuous monitoring of the transit egress. The first half of the transit window suffers from regular interruptions, which prevented us from monitoring the ingress. We perform a series of fits with different functional forms of the baseline models, including orbital phase polynomial orders from 4 to 7, overscan polynomial orders from 0 to 5 as well as the removal of the linear trend. The resulting transit shape (depth and duration) is found to be consistent at the 1-$\sigma$ level in each case. The adopted baseline model (in the BIC sense) remains the same as the one used for the global fit described above. We find no correlation of the transit signature in the photometry with any ancillary parameter (spectrum position, background or overscan values measured during data reduction). A zoom of the transit pattern is shown in Figure~\ref{fig:transp1}.

\subsection{July 2014 dataset}

The transit pattern detected in our July 2013 dataset prompted us to request new observations to confirm the transit signal, based on the radial-velocity period of 3.2357 days \citep{Dumusque:2012} and the transit centre computed from our STIS July 2013 observations. Nine orbits of HST DD time (PID 13927) were awarded to re-observe $\alpha$~Cen~B for 13.5 hours from 11:30 UT on 28 July 2014 to 01:35 UT on 29 July 2014.

\begin{figure*}
\begin{centering}
\includegraphics[width=\textwidth]{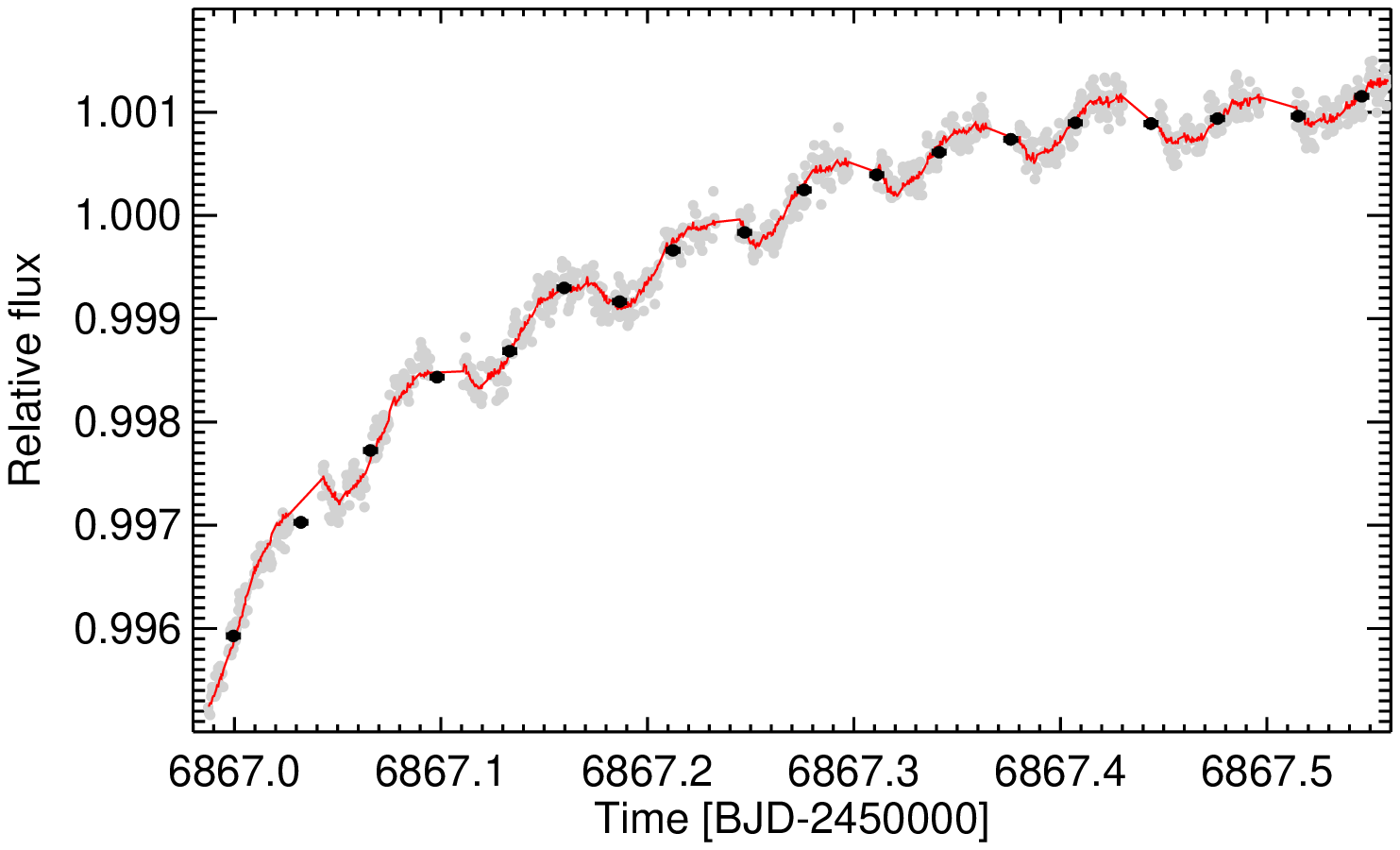}
\caption{{\bf July 2014 Raw HST/STIS photometry.} The single-baseline best-fit MCMC model is superimposed in red.
\label{fig:rawp2}}
\end{centering}
\end{figure*}

\begin{figure*}
\begin{centering}
\includegraphics[width=\textwidth]{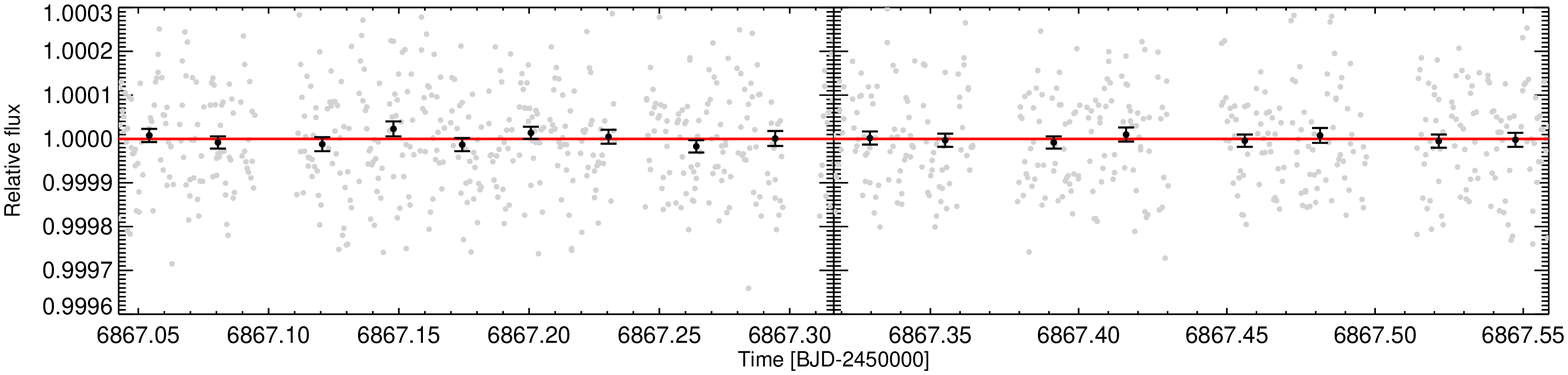}
\caption{{\bf July 2014 Detrended photometry \--- 2 segments.} Data are binned per 45 minutes. See text for details.
\label{fig:2segp2}}
\end{centering}
\end{figure*}

The data reduction and photometric analysis employed for this dataset are identical to the July 2013 observations. The raw photometry is shown in Figure~\ref{fig:rawp2}. We find that the first orbit of this visit exhibits a pronounced trend, likely due to the telescope settling on a new attitude. We do not notice such behaviour for the July 2013 dataset. We thus elect to discard the first 56 min (first orbit) of this visit. Ignoring the first orbit of a visit is a common procedure but is not found to be necessary for all STIS observations. The calibrated time-series for this second visit are shown in Figure~\ref{fig:2segp2}. We measure an RMS of 116 ppm and a red noise contribution of 20\%, in excellent agreement with the first visit's photometric properties, emphasising on the consistency of saturated spectroscopy with STIS.

We do not find any structure matching the transit pattern detected in the first dataset, while the photometric precision achieved for these 9 orbits would have enabled a clear detection of a 3.8-hr long, 100-ppm deep transit-like signature in the data at more than the 5-$\sigma$ level.

\section{No transit for Alpha~Cen~B\,b}
\label{sect:globanalysis}
In total, we observed $\alpha$~Cen~B for almost 40 hours in a consistent way with HST/STIS.  Both the July 2013 and 2014 datasets exhibit low-levels of correlated noise and a 6-s RMS of $\sim$115 ppm allowing us to confidently determine the a posteriori transit probability of the Earth-mass exoplanet $\alpha$~Cen~B\,b. We combine in the following both the 2013 and 2014 photometric datasets in an MCMC framework with the aim of determining to which extent our dataset rules out a transit of $\alpha$~Cen~B\,b with the published parameters of \citet{Dumusque:2012}.

We include both the published radial-velocity data and the STIS photometry described above as input data to the MCMC fit. We model the radial velocities following the appendix of \citet{Dumusque:2012}. We use the same functional form for the detrending model including the rotational activity, magnetic cycles and binary motion. The detrended radial-velocity datapoints are included in the MCMC fit and their errors include the model coefficient uncertainties. For the transit model, we compute the quadratic limb-darkening coefficients from a Kurucz stellar atmosphere model \citep{Kurucz:1979}. We first adopt a 5,250K, $\log g$=4.5 and [Fe/H]=0.2 model, which represents the closest match to $\alpha$~Cen~B's properties. We use the model output file to calculate the weighted averaged centre-to-limb intensity profile using STIS G750M's 6094 setting wavelength coverage. We then fit the profile with a two-parameter quadratic limb-darkening law \citep{Claret:2000}. We repeat the same procedure for the 6 neighbouring stellar atmosphere models in the 3-dimensional grid and use the uncertainties on $\alpha$~Cen~B's stellar properties to estimate individual errors on the limb-darkening parameters. We find the quadratic coefficients to be $u_1=0.504 \pm 0.010$ and $u_2=0.236\pm0.008$. The MCMC fit has the following jump parameters: the time of RV maximum $T_0$, the orbital period $P$, the parameter $K' = K\sqrt{1-e^2}P^{1/3}$, where $K$ is the radial-velocity semi-amplitude, the two parameters $\sqrt{e}\cos\omega$ and $\sqrt{e}\sin\omega$, the planet/star radius ratio $R_{P}/R_{\star}$ and the impact parameter $b$.
We also add the linear combinations $c_1=2u_1+u_2$ and $c_2=u_1-2u_2$ (where $u_1$ and $u_2$ are the quadratic coefficients determined above) as jump parameters to minimise the correlations of the resulting uncertainties \citep{Holman:2006}. We impose Gaussian priors on the limb-darkening coefficients.

We note that neglecting the 2014 dataset yields an orbital solution providing a satisfying fit to both the spectroscopic and photometric data, provided the orbit has a significant eccentricity ($e=0.54$) and a longitude of periastron $\omega \sim 270$ degrees. This orbital configuration allows $e \sin \omega$ to be significantly offset from 0, thus providing leverage to the fit in increasing the transit duration while leaving the transit $T_0$ almost unchanged. If this transit was caused by $\alpha$~Cen~B\,b, the transit depth would translate to a planetary size of $0.92\pm0.06$ Earth radii, which, combined to our revised mass measurement of $1.01\pm0.09$ Earth masses, results in a density that is 30\% larger than the Earth, similar to that of Mercury \citep{Hauck:2013}. Using specific dissipation function values of \citet{Goldreich:1966} to estimate circularisation timescales, we estimate eccentricity damping times of less than 100 Myr for a planet on such a large eccentricity orbit, which is significantly shorter than the age of 5-6 Gyr of the $\alpha$~Cen system \citep{Eggenberger:2004a}. While not impossible, it is unlikely that the eccentricity of such a small planet would be maintained at eccentricities $> 0.5$ as suggested by the transit signal duration. Combining both HST datasets rules out the link between the transit-like pattern seen in July 2013 and $\alpha$~Cen~B\,b published orbital parameters. Note however that the 2014 dataset probes the repeatability of the transit pattern within a 2-$\sigma$ window based on the published orbital period value. 

Our MCMC analysis also allows us to compute the {\it a posteriori} transit probability of $\alpha$ Cen B\,b based on the parameters of \citet{Dumusque:2012}. In other words we estimate to which level our combined {\it HST} dataset discards the transit of $\alpha$~Cen~B\,b as published. Using the published orbital parameters, the geometric {\it a priori} transit probability of $\alpha$~Cen~B\,b assuming a circular orbit is 9.5\%. For the purpose of this analysis, we put a normal prior on the planetary radius N(0.97,0.10$^2$), which is motivated by the $\sim$1 Earth-mass of the planet. We also include the radial-velocity data and consider only circular orbits. We find that for 3.4\% of the MCMC accepted steps, a transiting configuration is found requiring a $T_0\sim2455281.55$ and $P\sim3.2375$, at the edge of our observing window. These values have to be compared to the transit ephemerides of $T_0=2455281.24\pm0.17$ and $P=3.2357\pm0.0008$ computed from the radial-velocities alone, prior to the {\it HST} observations. The {\it a posteriori} transit probability is thus $0.034\times0.095=0.32$\%. Our observations thus allows to reduce the transit probability of $\alpha$~Cen~B\,b by a factor 30.

\section{Possible origin of the 2013 transit pattern}

We explore in the following the possibility that the July 2013 transit pattern is due to stellar variability,  instrumental systematics or cause by a background eclipsing binary. We do not find any temperature or HST orbital dependent parameter, nor X/Y spectral drifts to correlate with the transit pattern. The transit candidate duration of 3.8 hours is 2.4 times longer than the HST orbital period, making the transit pattern unlikely to be attributable to HST instrumental systematics. As the detector is consistently saturated during all of our observations, we also find it unlikely that saturation is the origin of the transit signal. Another possible explanation could be stellar variability. However, the duration of the transit candidate (3.8-hr) is not consistent with the stellar rotational period of 36.2 days \citep{DeWarf:2010}, to enable a spot (or group of spots) to come in and out of view. In such a case, star spots would change the overall observed flux level and produce transit-shape signals, as is the case for stars having fast rotational periods \citep[e.g.][]{Harrison:2012}. We finally find no sign of light contamination from the A component, which we specifically avoided during the preparation of both 2013 and 2014 visits. Our data reduction and analysis is straightforward, including only basic detrending variables related to the HST orbital phase and spectrum position. We find the transit pattern to be robust to the detrending approaches we employ.

We then estimate the probability that the observed transit of $91~\rm ppm$ is due to a background eclipsing binary (BEB). We first estimate up to which magnitude a diluted BEB can cause this signal by applying
\begin{equation}
\delta_{BEB} = \delta_{EB} \cdot \frac{F_{EB}}{F_{EB} + F_{target}}
= \delta_{EB} \cdot 2.512^{\Delta m},
\end{equation}
whereby we express the flux ratio via the magnitude difference $\Delta m$ between the BEB and the target star. We now allow all $\delta_{EB} \in (0,1]$ and find that the target star with $V=1.33$ can dilute BEBs of up to $V < 11.5$. Fainter BEBs cannot cause a deep enough signal, and only faint BEBs with significant transit depths or bright BEBs with grazing eclipses can cause the required depth. Secondary eclipses will further drastically reduce the fraction of reasonable systems. However, here we allow every scenario with $V < 11.5$ and thus significantly overestimate the total number of possible BEBs. Using the TRILEGAL \citep{Girardi:2005} galaxy model we find a distribution of around $1000$ star systems with $V < 11.5$ in $1$~sq.~deg. around the field of view, leading to $0.8\%$ probability to randomly have any of these in the $52 \times 2$~sq.~arcsec slit. The probability to have any kind of detached EB among all star systems is roughly $1\%$, shown by our simulations and confirmed by the results from Kepler \citep{Slawson:2011}. Thereby, we again overestimate the risk of such BEBs, since triple and higher-order systems additionally dilute any eclipse such that the magnitude limit for the BEB would be lower, leaving significantly less systems. The probability for the signal to be caused by a BEB is hence lower than $0.008\%$.

A possible remaining explanation for the observed signal is thus a planetary transit whose orbital period is significantly different from $\alpha$~Cen~B\,b's, as published in the discovery paper. We use our {\it HST} transit detection to constrain the orbital period of this putative planet and the observing window where another transit would happen, which would confirm the repeatability of the signal. The host star, Alpha Cen B, benefits from extensive characterisation. Both the stellar mass ($M_{\star}=0.934\pm0.006 M_{\odot}$) and radius ($R_{\star}=0.862\pm0.005 R_{\odot}$) are known to a remarkable precision \citep{Kervella:2003}. We thus know the stellar density, which can also be constrained by the transit shape \citep[e.g.][assuming a circular orbit.]{Seager:2003}. Combining our knowledge from the host star and the 2013 transit shape allows us to constrain the orbital period, impact parameter and eccentricity, albeit in a degenerate way. We employ our MCMC framework described above to explore what orbital periods/eccentricities/impact parameters are allowed by the {\it HST} 2013 transit lightcurve alone. We find that the STIS photometry yields an orbital period 2-$\sigma$ upper limit of 20.4 days (the median of the posterior being 12.4 days), with a modest impact parameter $\sim0.0-0.3$ and an eccentricity 2-$\sigma$ upper limit of 0.24.

The transit pattern detected in our HST/STIS data motivates further photometric follow-up observations of $\alpha$~Cen~B to confirm its repeatability. It is unlikely that the mass of an Earth-size planet orbiting $\alpha$~Cen~B with a significantly longer orbital period than $\alpha$~Cen~B\,b could be constrained in the near future. Since the radial-velocity semi-amplitude decreases with $1/P^{1/3}$, an Earth-mass planet on a e.g. 15-day orbit would yield a radial-velocity amplitude of 0.3 m/s, compared to K=0.5 m/s in the case of the 3.24-day $\alpha$~Cen~Bb. The photometric follow-up is equally complicated. Ground-based facilities do not have the ability to 1) continuously monitor the 2-3 weeks window and 2) to reach the 30 ppm per 3.8 hour precision necessary to detect a 90-ppm transit (assuming an Earth-like interior) at the 3-$\sigma$ level. The MOST satellite \citep{Walker:2003} performed several observations of bright stars. Considering the $\alpha$ Cen system, the transit signature would be diluted by a factor of three, in the combined light of the close A and B components, thus rendering its confirmation extremely difficult. $\alpha$~Cen~B's flux in {\it Spitzer}'s IRAC channels 1 and 2 is $\sim$371 and $\sim$220 Jy respectively, more than ten times IRAC's saturation levels. Our observing strategy and data reduction technique demonstrates that HST/STIS is able to detect 100-ppm transits at 5-$\sigma$ or more for stars that exceed the saturation level of the detector by a factor 5. The characterisation of small exoplanets orbiting stars as bright as $\alpha$~Cen~B will continue to pose significant challenges in the future unless next-generation of telescopes have the ability to employ successful observing strategies for highly-saturated stars.

\section*{Acknowledgments}

We are indebted to Shelly Meyett and Tricia Royle for their assistance in the planning and executing of these observations. We thank Tom Ayres for his help in preparing our phase 2 program, as well as Adrian Barker, Samantha Thompson and Julien de Wit for discussions. We thank the anonymous referee for a detailed and helpful review that improved the paper. We are grateful to STScI Director Matt Mountain for awarding Director's Discretionary Time for both programs. Based on observations made with the NASA/ESA Hubble Space Telescope, obtained at the Space Telescope Science Institute, which is operated by the Association of Universities for Research in Astronomy, Inc., under NASA contract NAS 5-26555. These observations are associated with program PID13180 and 13927. Support for this program was provided by NASA through a grant from the Space Telescope Science Institute, which is operated by the Association of Universities for Research in Astronomy, Inc., under NASA contract NAS 5-26555. This work has been carried out within the frame of the National Center for Competence in Research PlanetS supported by the Swiss National Science Fundation (SNSF). D.E., C.L., F.P., and S.U. acknowledge the financial support of the SNSF. A. T. is a Swiss National Science Foundation fellow under grant number P300P2-147773.

\footnotesize{
\bibliography{apj-jour,acenb_hst}
}

\bsp
\label{lastpage}

\end{document}